\def\year{2018}\relax
\documentclass[letterpaper]{article} 
\usepackage{aaai18}  
\usepackage{times}  
\usepackage{helvet}  
\usepackage{courier}  
\usepackage{url}  
\usepackage{graphicx}  
\expandafter\def\expandafter\UrlBreaks\expandafter{\UrlBreaks
  \do\a\do\b\do\c\do\d\do\e\do\f\do\g\do\h\do\i\do\j%
  \do\k\do\l\do\m\do\n\do\o\do\p\do\q\do\r\do\s\do\t%
  \do\u\do\v\do\w\do\x\do\y\do\z\do\A\do\B\do\C\do\D%
  \do\E\do\F\do\G\do\H\do\I\do\J\do\K\do\L\do\M\do\N%
  \do\O\do\P\do\Q\do\R\do\S\do\T\do\U\do\V\do\W\do\X%
  \do\Y\do\Z}
\usepackage{floatrow}
\usepackage{multirow}
\usepackage[inline]{enumitem}
\DeclareFloatFont{small}{\small}
\newcommand\blfootnote[1]{%
  \begingroup
  \renewcommand\thefootnote{}\footnote{#1}%
  \addtocounter{footnote}{-1}%
  \endgroup
}

\begin{document}
%
 \title{An Exploration of Verbatim Content Republishing by News Producers}

\author{Benjamin D. Horne and Sibel Adal{\i} \\
Rensselaer Polytechnic Institute, Troy, New York, USA\\
\{horneb, adalis\}@rpi.edu\\
}

\maketitle
\begin{abstract}
In today's news ecosystem, news sources emerge frequently and can vary widely in intent. This intent can range from benign to malicious, with many tactics being used to achieve their goals. One lesser studied tactic is content republishing, which can be used to make specific stories seem more important, create uncertainty around an event, or create a perception of credibility for unreliable news sources. In this paper, we take a first step in understanding this tactic by exploring verbatim content copying across 92 news producers of various characteristics. We find that content copying  occurs more frequently between like-audience sources (eg. alternative news, mainstream news, etc.), but there consistently exists sparse connections between these communities. We also find that despite articles being verbatim, the headlines are often changed. Specifically, we find that mainstream sources change more structural features, while alternative sources change many more content features, often changing the emotional tone and bias of the titles. We conclude that content republishing networks can help identify and label the intent of brand-new news sources using the tight-knit community they belong to. In addition, it is possible to use the network to find important content producers in each community, producers that are used to amplify messages of other sources, and producers that distort the messages of other sources. \blfootnote{Presented at NECO 2018, co-located with ICWSM 2018}
\end{abstract}

\section{Introduction} 
In the post-truth era~\cite{davies2016age}, the intent of news producers can vary widely. These motives range from upholding journalistic standards, to making money from clicks, to maliciously pushing an agenda~\cite{starbird2017examining}. The ease of establishing a news distribution entity in today's online ecosystem has created an increase in malicious and hyper-partisan news sources, as well as a more diverse set of malicious tactics and motives. Due to this increased diversity and complexity, recent work has focused on both characterizing and detecting malicious misinformation~\cite{horne2017just}, using title structures that encourage clicks~\cite{chakraborty2016stop}, hyper-partisan coverage~\cite{pennycook2018crowdsourcing}, and using bots to increase story visibility~\cite{shao2017spread} in news articles.   

One lesser studied tactic is content republishing. The most direct motivation for this is to increase the availability of specific stories to make it appear more widely known, accepted, discussed, current, and mainstream. It is also possible that sources produce near identical copies of information already receiving a lot of attention to generate revenue through clicks. It has been hypothesized that malicious news producers enforce political agendas by not only spreading false or misleading information, but also by causing uncertainty around an event or political stance. As a separate method, malicious sources can create uncertainty by publishing credible news stories along side misleading ones. This notion has been informally explored in~\cite{lytvynenko}, where a well-known conspiracy news source, Infowars, is shown to have copied many true articles from credible sources without attribution. This notion is further supported by the unreliable context indicators identified in~\cite{zhang2018structured}. Content mixing can happen at many different levels of 
granularity, with or without attribution, and with different intent. To be able to develop
algorithms to identify the different roles source can play in the news ecosystem, we need the 
conduct a first analysis of content republishing methods from a wide-range of sources over a 
span of time. This is the focus of this paper.


As a first step in understanding this tactic, we explore verbatim content copying across 92 news and media producers of different characteristics over a span of 4 months. Specifically, we ask three questions:

\begin{itemize}
    \item \textbf{Q1}: What are the different contexts under which verbatim content republishing occurs?
    \item \textbf{Q2}: Does attribution behavior differ between alternative and mainstream news producers?
    \item \textbf{Q3}: Of those sources that republish articles (correctly or incorrectly), do they change the title, and if so, how?
    \item \textbf{Q4}: Which copies of the same information get more attention in social media?
\end{itemize}

We employ a mix of quantitative and qualitative approaches to explore the data. First, we build content similarity networks to assess the community structure among news publishers (see Figure~\ref{attrib_nets}). Further, the network structure allows us to explore consistency in republishing behavior over time and what news sources are commonly re-publishers or producers. Next, we perform a manual qualitative analysis on republished article pairs to gain better insight to our network methodology, as well as gain a better understanding of what types of attribution exist between news source pairs. Lastly, we explore how news producers change the headlines of republished articles by using a large set of well-studied natural language features and basic statistical analysis.

Our study provides some interesting insights into the news ecosystem. The content copying 
occurs more frequently between like-audience sources (eg. alternative news, mainstream news, etc.), but there consistently exists sparse connections between these communities. 
We find that the majority of republished articles provide some level of citation or are written by the same author for multiple news producers. Further, we find that 58\% of the republished articles change the headline in some way. Of those news sources that change headlines, we find that mainstream sources change more structural features, while alternative sources change many more content features, often changing the emotional tone and bias of the titles. 
In short, we conclude that 
the content copying network makes it possible to identify communities with respect to different
types of labels. Furthermore, it is possible to use the network to find frequent and prominent content producers in each community, producers that are used to amplify message of some sources, as well as producers that aim to distort the messages of other sources. Hence, it is worthwhile to 
study the content copy network as a whole instead of concentrating on individual sources, not only to understand the specific role sources play in the ecosystem, but also to quantify the cumulative
impact a community of sources have on the information consumers.

\section{Related Work}
Overall, there has been very little work on content copying and attribution among news producers, specifically among alternative news producers. Most related is work on text similarity methods for identifying text reuse in news. Pal and Gilliam propose an approach to linking similar news articles in a cross language environment using a similarity ranking model~\cite{pal2013set}. They show reasonable accuracy in identifying matches. Palkovskii and Belov provide a similar study~\cite{palkovskii2011using}. More generally, Ioannou et al. propose a method to detect near duplicate text resources for grouping data on the semantic and social web~\cite{ioannou2010efficient}. The most obvious use case of this method is for online news. Potthast et al. proposes a technological remedy to synthesize original pieces of text without text reuse to prevent ancillary copyright issues during web search~\cite{potthast2018plan}. Clough et al. build a corpus for the study of journalistic text reuse called METER~\cite{clough2002building}~\cite{gaizauskas2001meter}. METER identifies varying degrees of text reusing, including verbatim, rewrite, and new. This corpus has been used to build several text reuse detection methods for news~\cite{bar2012text}~\cite{adeel2012detecting}. While all these works explore online text reuse at some level, their goal is to create new methodologies for detecting and mitigating online text reuse. The goal in our work is not to create a new method for identifying text reuse, but to explore text reuse behavior in the current news ecosystem, including among malicious and hyper-partisan news sources. Further, our exploration only focuses on verbatim content copying, rather than various levels of reuse. 

Also related is work on news republishing in journalism and the social sciences. These works have mostly been focused on copyright law for online news and news aggregators~\cite{quinn2014associated}~\cite{rowland2003whose}. In this study, we choose to not address copyright issues as they may be diverse among pairs of news sources, and this is information not directly available to us.

\begin{table*}
\begin{center}
\fontsize{8}{8.5}\selectfont
\hspace*{-0.0in}\begin{tabular}{|c|c|c|c|c|}
\hline
\multicolumn{5}{|c|}{\textbf{Sources in Dataset}}\\
\hline
AP   & Freedom Daily   & Observer   & Duran & Drudge Report \\
Activist Post  & Freedom Outpost  & Occupy Democrats  & Fiscal Times & Young Conservatives  \\
Addicting Info  & FrontPage Mag   & PBS   & Gateway Pundit & Yahoo News \\
Alternative Media Syndicate   & Fusion  & Palmer Report   & The Guardian & Xinhua \\
BBC  & Glossy News   & Politicus USA  & The Hill & World News Politics\\
Bipartisan Report  & Hang the Bankers  & Prntly  & Huffington Post & Waking Times   \\
Breitbart   & Humor Times  & RT  & The Inquisitr & Daily Beast  \\
Business Insider  & Infowars  & The Real Strategy  & New York Times   & Newslo \\
BuzzFeed  & Intellihub  & Real News Right Now  & The Political Insider & Fox News  \\
CBS News  & Investors Biz Daily  & RedState   & Truthfeed  & Vox \\
CNBC    & Liberty Writers   & Salon   & The Right Scoop  & D.C. Clothesline\\
CNN    & Media Matters   & Shareblue   & The Shovel  & NewsBusters \\
CNS News    & MotherJones   & Slate   & The Spoof  & Faking News \\
Conservative Tribune    & NODISINFO & Talking Points Memo  & TheBlaze & Veterans Today \\
Counter Current  & NPR   & The Atlantic   & ThinkProgress & Conservative TreeHouse  \\
Daily Buzz Live  & National Review  & The Beaverton   & True Pundit & NewsBiscuit  \\
Daily Kos  & Natural News  & Borowitz Report  & Washington Examiner &   \\
Daily Mail & New York Daily  & Burrard Street Journal  & USA Politics Now  &\\
Daily Stormer  & New York Post  & The Chaser  & USA Today  &\\
\hline
\end{tabular}
\caption{Sources in NELA2017 data set. We use all sources in our analysis, but only find that 67 of the 92 sources have had at least 1 article highly similar to another source's article (copied or copied from) between April 2017 and July 2017}\label{sources}
\end{center}
\end{table*}

\section{Data}
In order to adequately explore content copying and attribution in the news, we need a large and wide range sample of news producers. To this end, we extract data from the NELA2017 data set~\cite{horne2018sampling}. The NELA2017 data set is a near complete set of news articles from 92 news producers between April 2017 and October 2017. The data set contains a wide range of news producers, including mainstream news, political blogs, satire websites, and many alternative news sources that have published misinformation in the past or have relatively unknown veracity. These news sources can be found in Table~\ref{sources}. The NELA2017 data set provides the following information for each article:

\begin{itemize}
    \item the article content and title
    \item the author of article according to the article webpage
    \item a UTC time stamp of publication
    \item the link to the original article if it still exists online.
    \item the html of the original article
\end{itemize}

For this study, we extract all articles from all 92 sources between April 7th 2017 and July 14th 2017, amounting to over 54k articles. We use the, article content, the UTC timestamp, and the author data for our study.

\section{Methodology}
Using this large and near complete data set, we extract highly similar articles. To do this, we first divide the data set into 2 week intervals. We do this divide for several reasons:
\begin{enumerate*}
  \item It may be more insightful to see how copying behavior changes over time. Do news producers copy from the same sources in every time slice or is the behavior inconsistent?
  \item Analyzing a smaller time frame both decreases run time and memory constraints.
  \item We assume most articles that are copied are copied within a short time frame.
\end{enumerate*}

In each 2-week divide, we create a Term-Frequency Inverse-Document-Frequency (TFIDF) matrix treating each article's body-text as a document. Then we compute the cosine similarity between all pairs of article vectors. We extract article pairs that have cosine similarities greater than 0.90 and are published by differing sources. A cosine similarity above 0.90 means the articles are almost word-for-word copies of each other. TFIDF is a standard technique in information retrieval to determine the importance of a word in a text document or collection of documents~\cite{ramos2003using}. It is commonly used to find the most similar documents in a corpus~\cite{tata2007estimating}.

For each extracted article pair, we compare the time-stamps to find which source published the article first. This time-stamp comparison is then used to create a weighted directed graph of all sources in the two week time-frame. Each edge $A\rightarrow B$ from a source $A$ to a source $B$ indicates that $A$ copied from $B$. The weight of the edge is determined by how many articles were copied during the the two-week time frame. Thus, a source's weighted in-degree is the total number of articles copied from that source. Notice, since this is a pair-wise analysis, there may be
redundant links if the same story is copied by many sources.
For example, if several sources copy a story from The Associate Press, the
network will not only point to The Associate Press, but also to the sources
that published that story earlier than another source. In our qualitative analysis, we do not find this redundancy to be the case overall, but more frequently among mainstream and news-wire sources. This methodology is very simple, but effective for our task. Since we want to explore near verbatim copies, we do not use a semantically-aware similarity method. This analysis is left for future work.

In total, 67 of our 92 sources had at least 1 article highly similar to another source's article (copied or copied from). 

\begin{figure*}[h]
\begin{center}
\hspace*{-0.3in}\begin{tabular}{cc}
\small{(a) Top 10 - Highest In-Degree (total over all time frames)} & \small{(b) Top 10 - Highest Out-Degree (total over all time frames)}\\
\includegraphics[width=245pt,keepaspectratio=true]{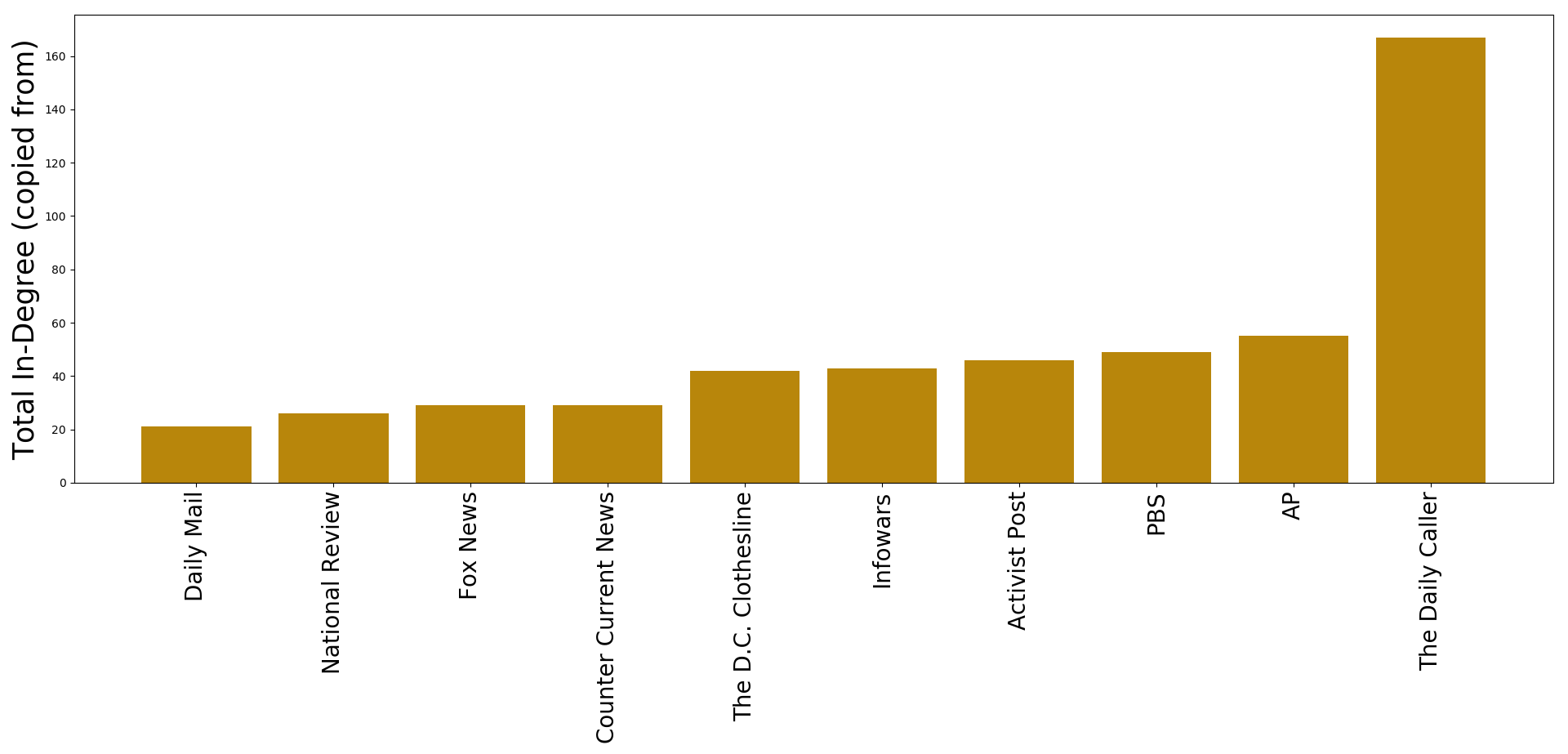}&
\includegraphics[width=245pt,keepaspectratio=true]{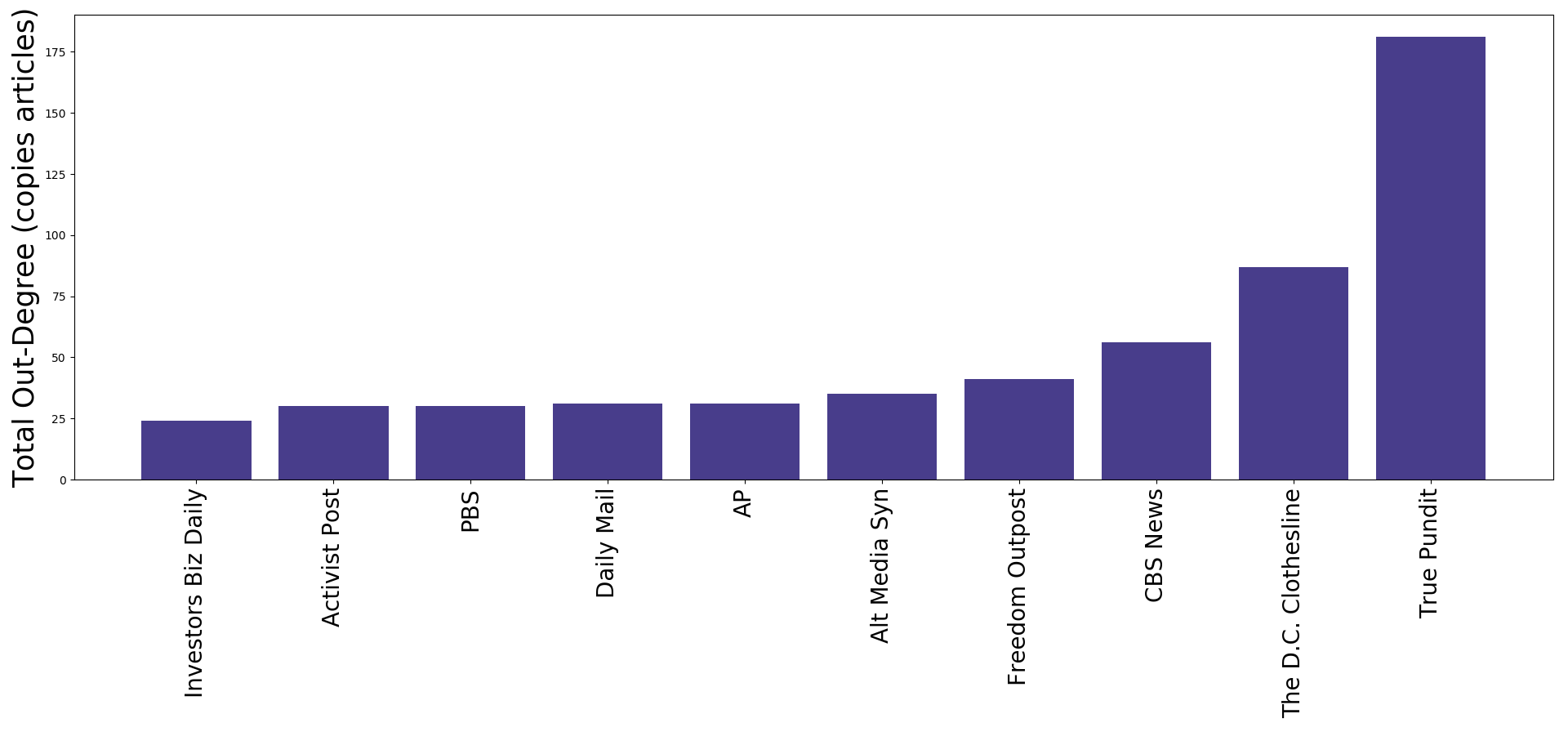}\\
\small{(c) Top 10 - In-Degree Centrality (averaged over time slices)} & \small{(d) Top 10 - Betweenness Centrality (averaged over time slices)}\\
\includegraphics[width=245pt,keepaspectratio=true]{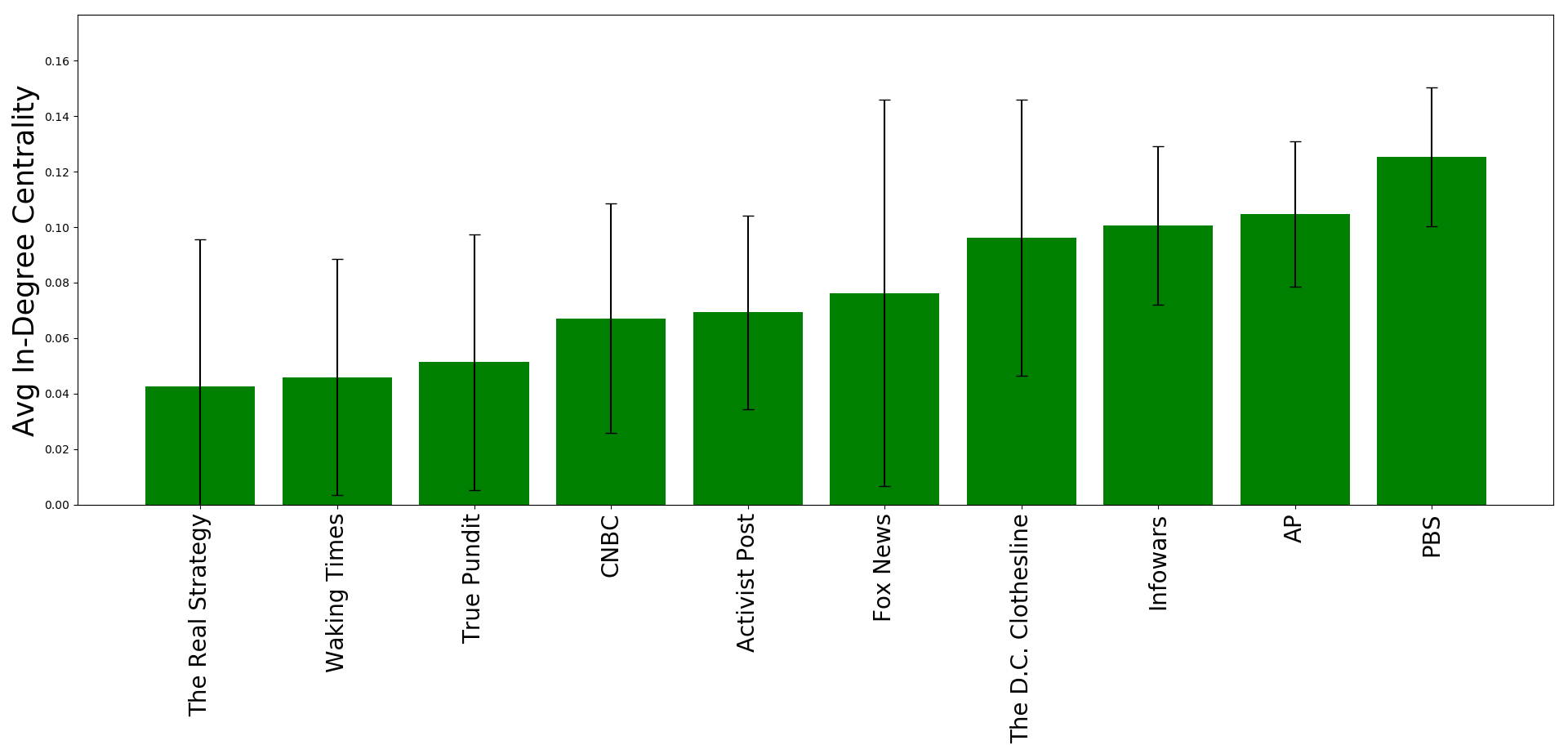}&
\includegraphics[width=245pt,keepaspectratio=true]{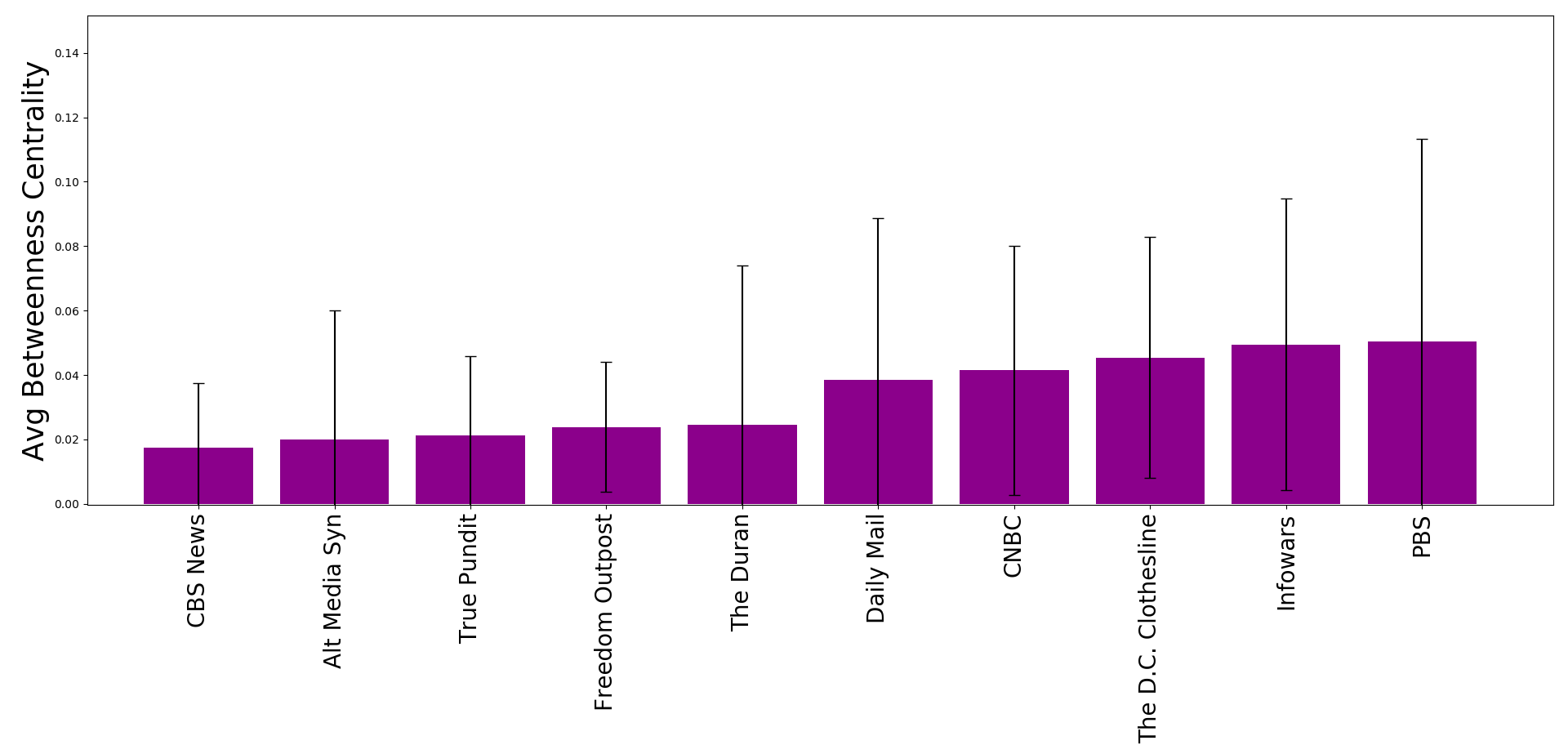}\\
\end{tabular}
\caption{Feature distributions across different articles from specific
  sources (Bars in Figures (c) and (d) represent the variance across time slices) \label{meta}}
\end{center}
\end{figure*}

\section{Results}
\begin{figure*}[ht] 
\centering
\hspace*{-0.8in}\begin{tabular}{cc}
\small{(a) Colors = Communities by Modularity} & \small{(b) Green = Alternative, Blue = Mainstream, Red=Satire/Unknown}\\
\includegraphics[width=10.8cm]{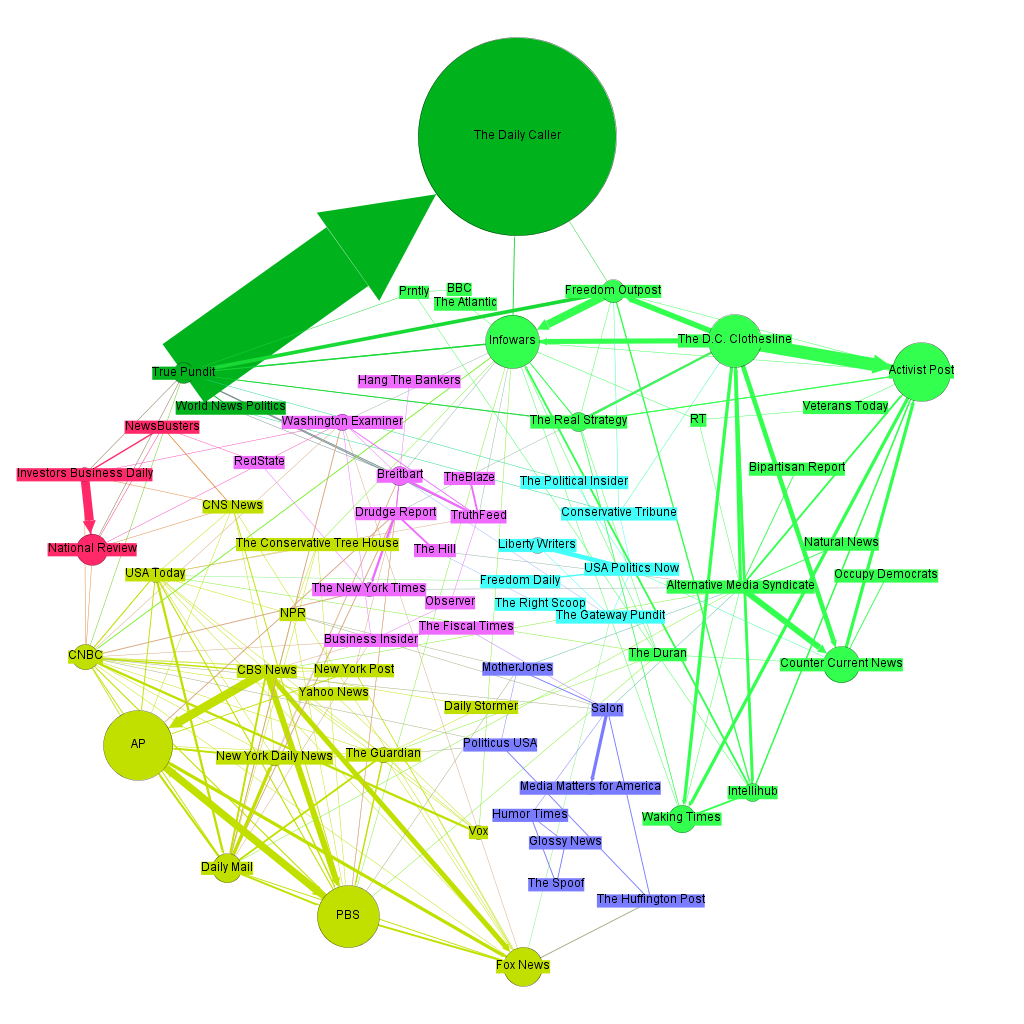}&
\includegraphics[width=10.8cm]{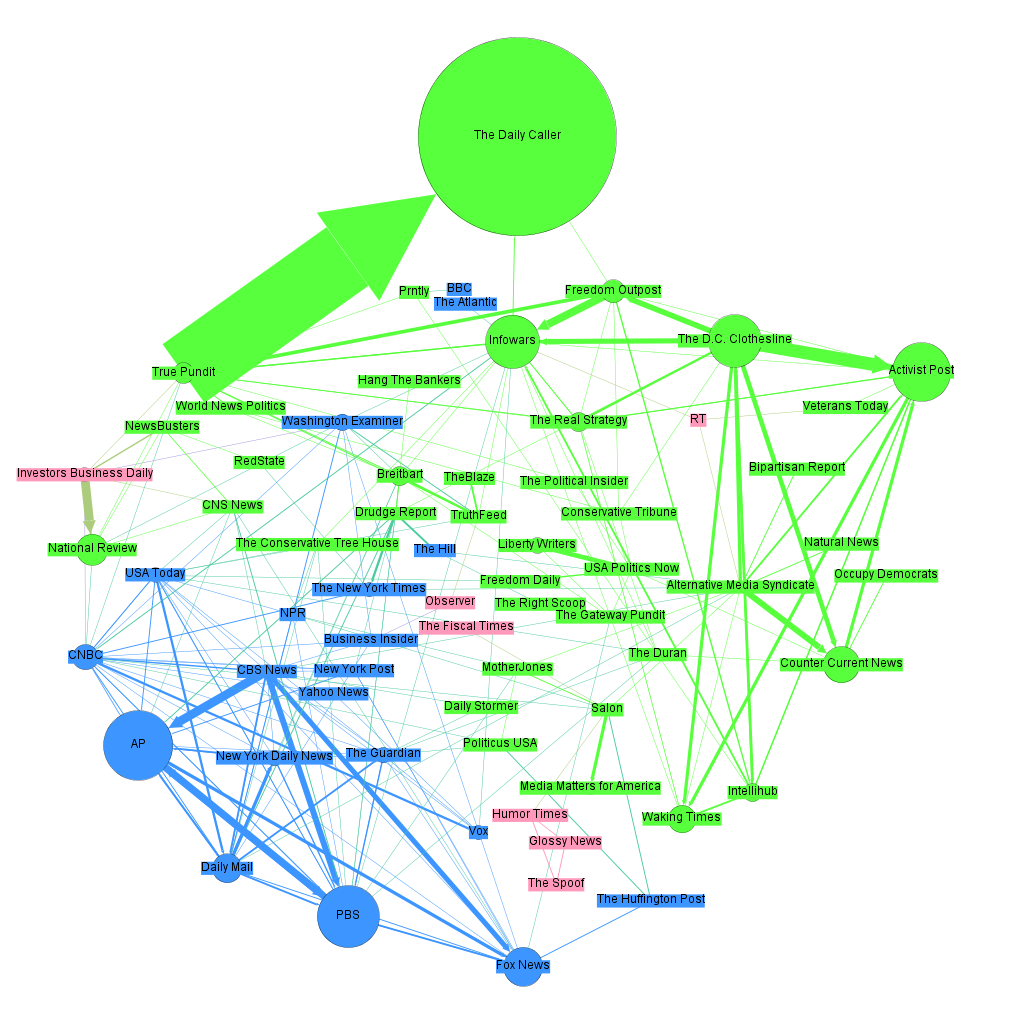}\\
\small{(c) Orange = Has Published Fake, Blue = Has Not/Unknown, Yellow = Satire} & \small{(d) Red = Right, Blue = Left, Yellow = Neutral/Unknown}\\
\includegraphics[width=10.8cm]{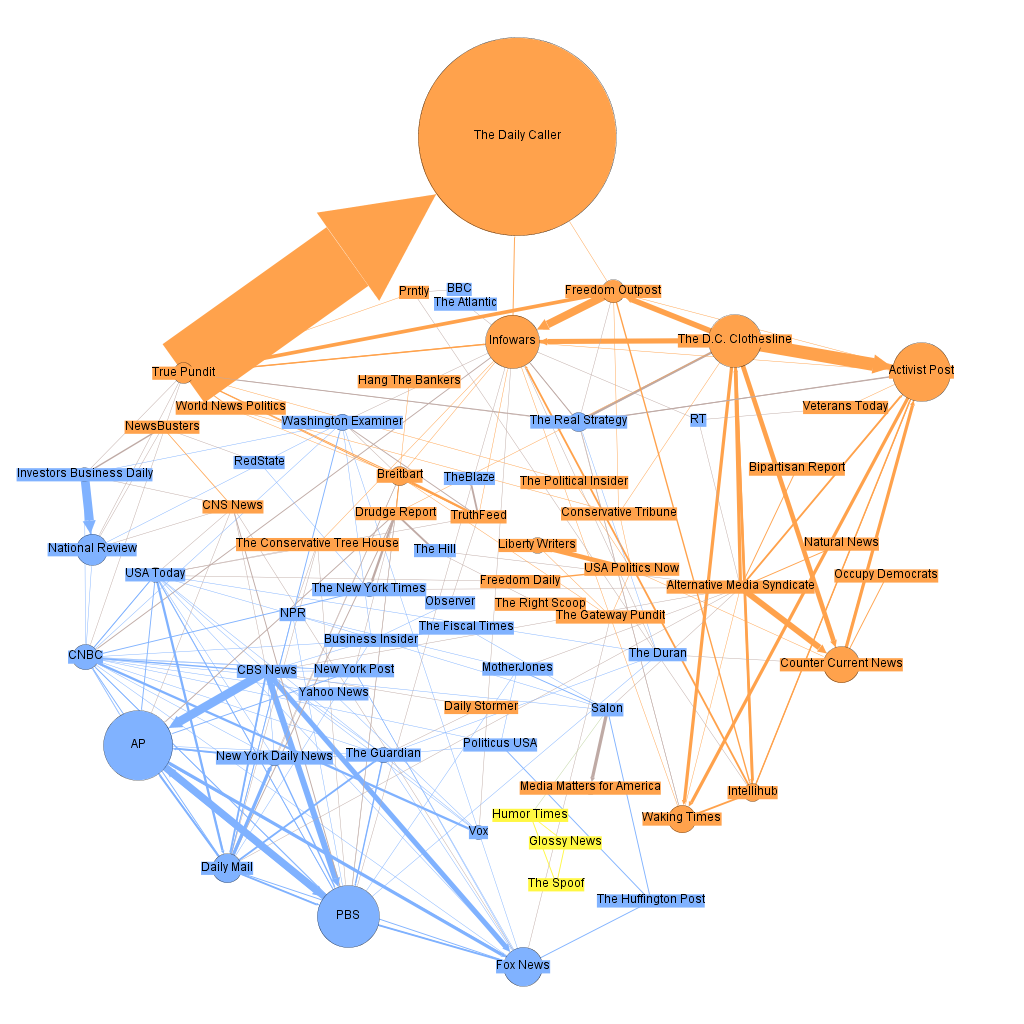}&
\includegraphics[width=10.8cm]{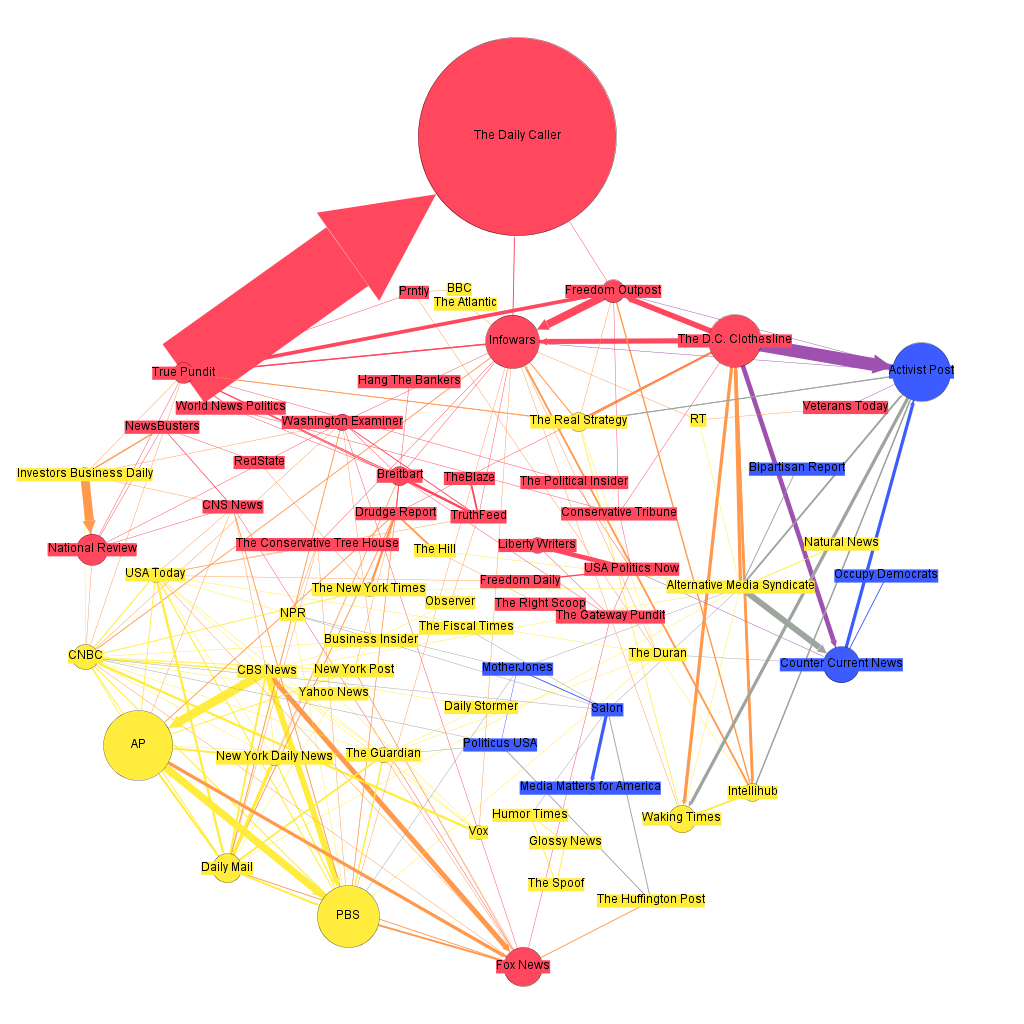}
\end{tabular}
 \vspace*{-0.2in}  \caption{Article similarity graphs during all 6 two-week periods. The weighted in-degree is the number of articles copied from a source. The weight is indicated by the size of the arrow. The in-degree of a source is shown by the size of the node.} 
\label{attrib_nets}
\end{figure*}

In Figure~\ref{attrib_nets}, we display the combined network structure by adding edges and weights across all time slices. In this figure, node size represents in-degree, arrow size represents edge weight, and node color represents a node category. Specifically, we categorize (and color) sources using 4 different criteria: 
\begin{enumerate}
    \item Community membership with respect to modularity, a commonly used measure of division in networks (Figure~\ref{attrib_nets}a)
    \item Mainstream or alternative news source labels from the lists in ~\cite{pennycook2018crowdsourcing} and ~\cite{potthast2017stylometric}. (Figure~\ref{attrib_nets}b) 
    \item Reliable or Unreliable labels based on online fact-checkers such as Snopes and Politifact. Specifically, if the source has published a completely false article in the past according to these fact-checkers, we consider the source unreliable. (Figure~\ref{attrib_nets}c) 
    \item Hyperpartisan right and left leaning source labels from~\cite{pennycook2018crowdsourcing}. If a source is not in the lexicon, we label it as neutral or unknown (Figure~\ref{attrib_nets}d). Note that a very small number of sources in our data that were likely hyperpartisan but were left out of this list. We kept them as neutral.
\end{enumerate}

Keep in mind, copying news articles can be done legitimately or illegitimately depending on copyright, attribution, and distribution permissions.

\subsection{Network Analysis}
For simplicity, we will show all our analysis in the combined network of all time slices in Figure~\ref{attrib_nets}. As we can see in Figure~\ref{attrib_nets}a, there are multiple components and clear communities in the combined graph. While we do not show in this paper, we are able to find the same communities in the graphs corresponding to each of the six time slices as well as the combined graph. In particular, we find two primary communities that can be roughly labeled as mainstream media and alternative media if we correlate them with the labels in Figure~\ref{attrib_nets}b. For the most part, these two communities are sparsely connected, with only a few links between them. In addition to these communities, we see smaller completely disconnected communities of satire media and ideologically-partisan media. Overall, these community structures are very similar to that of the news ecosystem on Twitter~\cite{starbird2017examining}, where alternative news sources form tight-knit communities with few connections to mainstream news. While these content copying communities are sparsely connected, there are several interesting paths between them. There exists a path from the mainstream community to the alternative community through The Gateway Pundit copying articles from Fox News. Similarly, in multiple time slices, we can see Infowars copying articles from PBS and CNBC. Similarly, there exists multiple paths from the mainstream media to the alternative media through Daily Mail. This behavior could reflect the hypothesis that maliciously false or conspiracy news sources mix real news stories into their publication to gain credibility among their audience. Further, many alternative sources copy from Infowars, allowing genuine news stories to be mixed with conspiracy in multiple sources. 


To get a bigger picture view, we rank nodes by in-degree (Figure~\ref{meta}a), out-degree (Figure~\ref{meta}b), in-degree  (Figure~\ref{meta}c), and betweenness centrality (Figure~\ref{meta}d) measures with respect to the combined network. These rankings show both expected and unexpected results. First, we see that The Associated Press (AP) has one of the highest in-degrees and in-degree centralities. This result is expected as AP primarily serves as a news wire service for other publishers and is well-established as a non-profit news agency. In addition to AP, Public Broadcasting Service (PBS) has high in-degree and in-degree centrality. PBS has a similar background to AP, being a well-established non-profit news agency. Looking at the networks in Figure~\ref{attrib_nets}, we see both AP and PBS as central nodes in the mainstream media communities. More interestingly, several alternative news sources have very high and consistent in-degree and in-degree centrality, including Infowars, The D.C. Clothesline, and Activist Post. All of these self-proclaimed alternative news sources have published well-known conspiracies and fake news stories in the past and appear to be a source of original content to other alternative new sources. 

We can also see that many of the nodes that are copied from the most also copy the most. As a result, many high in-degree nodes also have high betweenness centrality (Figure~\ref{meta}d). Betweenness appears to vary a great deal between time slices, but some nodes appear consistently across all time slices:  
The D.C. Clothesline, Infowars, and PBS. 

When looking at out-degree (those who copied articles the most), we see that True Pundit copies many more articles than any other source. This behavior is clearly seen in all time slices, where True Pundit has a heavy weighted edge to The Daily Caller. In total, True Pundit copies over 160 articles directly from The Daily Caller. The D.C. Clothesline, Alternative Media Syndicate, and Freedom Outpost copy many articles from other alternative sources. While, CBS and PBS report articles from other mainstream sources.

\subsection{Qualitative Analysis of Attribution}
In order to gain further insight into these copying networks, we perform a manual qualitative analysis of the types of copying found the network. We categorize types of copying into three categories: \begin{enumerate*}\item proper attribution, different title \item same author, different source \item no attribution \end{enumerate*}.

\paragraph{Proper Attribution, Different Title.}
Many sources publish full, verbatim articles using proper citation. For example, many sources, both mainstream and alternative, publish full articles from The Associated Press (AP), but provide clear citations. As previously mentioned, this finding is expected as AP often acts as a news wire service for other news producers. Sources such as PBS and CBS News are similarly cited, from both mainstream and alternative sources. While the content is typically verbatim in these cases, the titles can be different. We explore these title differences in Section~\ref{title_analysis}.

We also often see citation when alternative news sources copy from each other, but the titles are often left unchanged. For example, in all 160 articles copied by True Pundit from The Daily Caller, The Daily Caller is cited and the titles remain unchanged. This citation, like the citations of AP, seems to follow the producers content republishing protocol. Specifically, The Daily Caller writes ``Content created by The Daily Caller News Foundation is available without charge to any eligible news publisher that can provide a large audience" at the end of each article. True Pundit provides this information and licensing contact information for The Daily Caller at the end of their articles. Infowars similarly takes articles from the Daily Caller using the same attribution method, but to a much lesser degree than True Pundit. 

Some other pairs that cite the articles original source include: The Gateway Pundit taking from Fox News, CNBC taking from USA Today, Business Insider taking from Washington Examiner, Truthfeed taking from Fox News, Truthfeed taking from The Blaze, and Truthfeed taking from Breitbart. Notice, while these news sources cite where they directly copied information from, they may not all be legal copies. For example, when Truthfeed cites Breitbart, they simple state "Breitbart says:" above the fully copied article. We do not have knowledge of legal agreements or licensing of content between these sources. Simply indicating where the article came from still may be breaking proper attribution, but this is information not available to us. The Gateway Pundit taking from Fox News could be a similar case. As well as many of the news sources in the alternative media communities.

\paragraph{Same Author, Different Source.}
Frequently, we find that identical articles are written by the same author for different publishers, typically without citing the other publishers. There are many examples of this behavior:
\begin{itemize}
    \item Luke Rosiak writes identical articles for The Daily Caller, Infowars, and The Real Strategy.
    \item Alex Pietrowski and Makia Freeman write identical articles for The Waking Times and Activist Post.
    \item Cydney Hargis writes identical articles for Salon and Media Matters for America
    \item Michelle Malkin writes identical articles for National Review and CNS News.
    \item Jay Syrmopoulos writes identical articles for The D.C. Clothesline and Activist Post
    \item Rodger Freed writes identical articles for The Spoof, Humor Times, and Glossy News (all satire news sources).
\end{itemize}

In another example, a series of stories about a “George Soros backed Trump resistance fund” are published verbatim on both Infowars and Fox News, all written by Joe Schoffstal. Each article does not have clear attribution to one or the other source, despite being exact copies, and each article was written on Infowars days prior to its publication on Fox News. This example is particularly surprising as Fox News captures a wide, mainstream audience and Infowars is a well-known conspiracy and psuedo-science source. This creates a direct path  in the opposite direction of what we would expect (mainstream publishing an article from an alternative source).

Again, it is important to note, while these websites claim they are by the same author, some of them may not be legitimate. In many cases, it is obvious that the author writes for multiple sources through external website biographies, Twitter accounts, and the like, but in many other cases there is not clear indication that they write for multiple sources. For example, The D.C. Clothesline has many authors that contribute to other publications, but there is no clear indication these authors actually write for The D.C. Clothesline (no biographical information, list of other sources they contribute to, etc.) Hence, while the sources indicate the journalist who wrote the articles, even making them look like legitimate contributors to the website, this may not be proper attribution. However, without contacting each author in the data set, it is hard to obtain this knowledge. One very odd case of the same author writing in two sources is when The D.C. Clothesline publishes articles written by Activist Post authors, as the Activist Post often has completely opposing views as The D.C Clothesline (Activist Post is typically left leaning, while The D.C. Clothesline is typically right leaning). This leads us to believe The D.C. Clothesline may be faking the author postings.

\paragraph{No Attribution.}
Lastly, while surprisingly less common than the other two types of content copying, a still prevalent occurrence is copying without any citation, attribution, or author recognition. Veterans Today, Infowars, and Alternative Media Syndicate copy multiple articles verbatim from Russia Today (RT, a Russian government-controlled news site) with no citation. Similar behavior by Infowars has been pointed out by Jane Lytvynenko (2017). USA Politics Now copies articles from Liberty Writers, The Gateway Pundit, and Freedom Daily without citation (USA Politics Now was shut down in July 2017).

We do see a few outliers that do not fit into these three categories. For example, in Figure~\ref{attrib_nets}, The Daily Stormer appears to be a highly copied source, however this is due to publishing word for word speeches from U.S. President Donald Trump. The Daily Stormer published these speeches almost a full day earlier than the mainstream sources. In this case, it is highly unlikely a mainstream source actually copied the speech from The Daily Stormer. The Daily Stormer is known more as a hate-group than a news agency. Political speeches do not show up often in our data set, but do show a potential limitation in our network analysis.


\subsection{Facebook Engagement}
While our network analysis alone helps shed light on copying in the news ecosystem, it does not tell us much about the engagement of copied articles. To better understand this, we extract both the Facebook shares and Facebook reactions for each copied article pair in our 12-week time-frame. This information is available for all articles in the NELA2017 data set~\cite{horne2018sampling}. With this information, we color nodes based on the median engagement of all matched articles by a source, where the darker shade means more engagement. In Figure~\ref{attrib_nets2}a and~\ref{attrib_nets2}b, we show this for Facebook shares and reactions separately, although they are highly correlated. For the most part, we find that sources who produce the original article are more highly shared and reacted to. For example, The Daily Caller is much more highly shared than True Pundit, Occupy Democrats, Natural News, and the Bipartisan Report are much more highly shared than Alternative Media Syndicate, and Infowars is more highly shared than The D.C. Clothesline and Freedom Outpost. One possible way to explain this would be that highly shared articles tend to get copied over to other alternative sources to increase their visibility and potential credibility. The exception to this finding is in the mainstream community. We find that more newswire-like services such as AP and PBS are not as highly shared as those that copy from them. One may think this is due to some additional commentary added by those who copy newswire articles, but since we are only looking at verbatim copies, this cannot be the case. It is likely the case that sources such as The New York Times, Vox, NPR, The Huffington Post, and The Guardian simply have a larger audience (particularly on social media) than AP and PBS. 


\subsection{Headline Analysis}~\label{title_analysis}
\begin{figure*}[h]
\begin{center}
\hspace*{-0.3in}\begin{tabular}{cc}
\small{(a) Most Titles Changed} & \small{(b) Change Titles by Most}\\
\includegraphics[width=245pt,keepaspectratio=true]{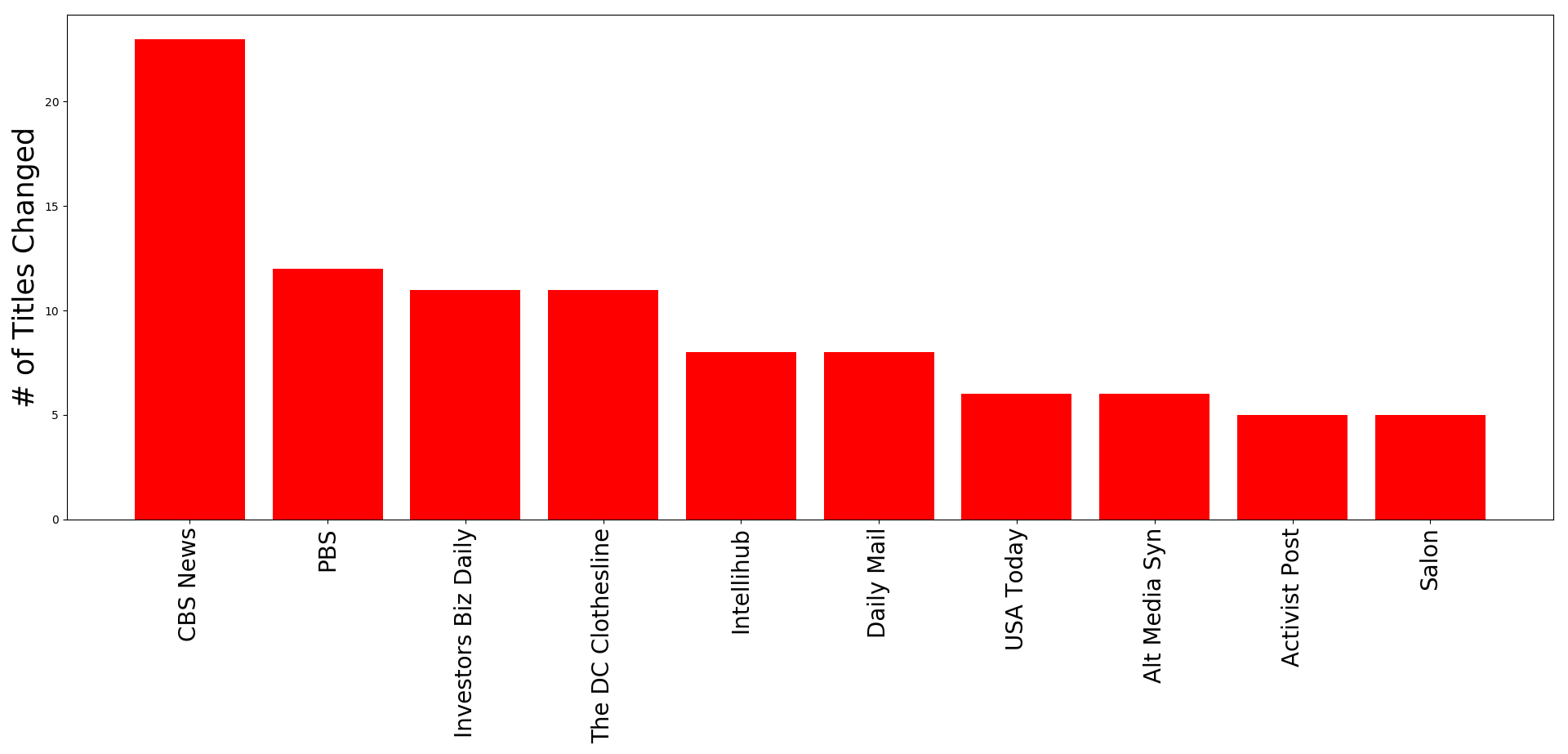}&
\includegraphics[width=245pt,keepaspectratio=true]{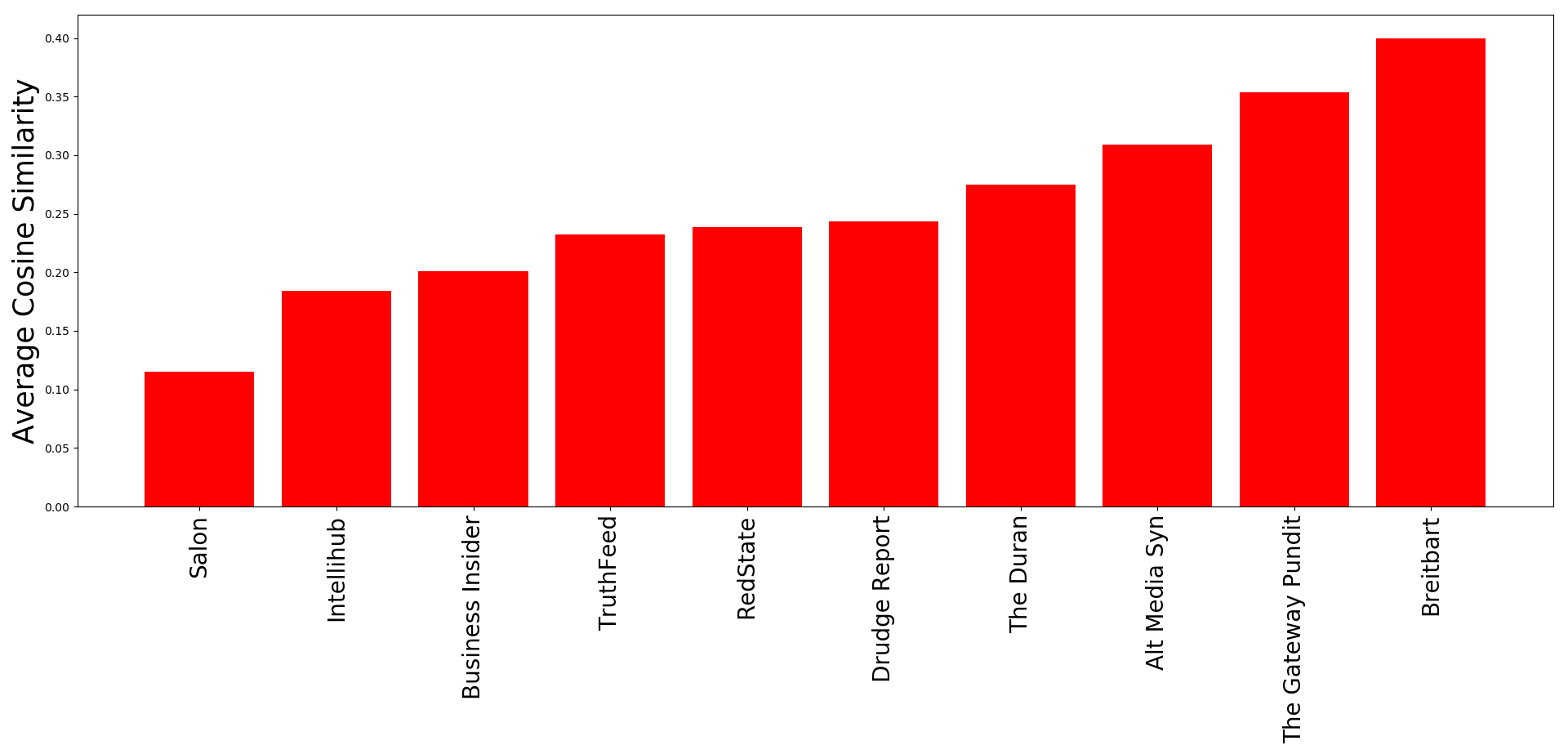}\\
\end{tabular}
\caption{(a) Top 10 news producers based on how many titles they change from original (b) Top 10 news producers based on how much they change titles (using 1 - cosine similarity).\label{title_top}}
\end{center}
\end{figure*}

As discussed in the previous section, many of the articles copied verbatim have different titles. It is well studied that news headlines can impact how consumers perceive, interpret, and share information. Ecker et al. show that misleading news headlines can affirm secondary information rather than the primary information of the article~\cite{ecker2014effects}. Surber and Schroeder show that, in general, titles impact the recall of information~\cite{surber2007effect}. While both Reis et al. and Piotrkowicz et al. show the importance of news headlines in determining news popularity~\cite{reis2015breaking}~\cite{piotrkowicz2017headlines}. In the case of highly similar articles, the headline becomes a focal point for misleading information. Thus, orthogonal to our main research question, we ask: of those sources that copy articles (correctly or incorrectly), do they change the title, and if so, how? To perform this analysis, we compute the cosine similarity between all copied article titles and explore the differences.  We find that $58.57\%$ of copied articles have titles that differ from the original by a cosine distance more than $0.10$. In Figure~\ref{title_top}a, we show what sources change the most titles. In Figure~\ref{title_top}b, we show the sources that change the title by the most when they change titles (based on cosine distance). These two graphs show different sets of news producers, where sources such as CBS News and PBS change many titles, but sources such as Breitbart and The Gateway Pundit change titles by the most. The types of changes made between title can vary greatly. This variance is clear looking at the following examples:\\
\textbf{Example 1:}
\begin{enumerate}
\item \underline{ORIGINAL}: (Breitbart) EPA Chief Scott Pruitt Calls for Exit of Paris Climate Agreement
\item  \underline{COPY}: (Truthfeed) BREAKING Trumps EPA Chief Makes Dramatic Announcement Liberals Crying
\end{enumerate}
\textbf{Example 2:}
\begin{enumerate}
\item  \underline{ORIGINAL}: (AP) Absences fitness atmosphere - new ways to track schools
\item  \underline{COPY}: (PBS) Beyond test scores here are new ways states are tracking school success
\end{enumerate}
\textbf{Example 3:}
\begin{enumerate}
\item  \underline{ORIGINAL}: (USA Today) Trey Gowdy new Oversight Committee chair plans to deemphasize Russia investigation
\item  \underline{COPY}: (Freedom Daily) BREAKING Trey Gowdy Exposes Massive Scam By Hillary And Obama
\end{enumerate}
\textbf{Example 4:}
\begin{enumerate}
\item  \underline{ORIGINAL}: (NPR) Republicans Now Control Obamacare - Will Your Coverage Change
\item  \underline{COPY}: (Salon) Death by 1000 cuts How Republicans can still alter your health coverage
\end{enumerate}

\begin{figure*}[ht] 
\centering
\hspace*{-0.8in}\begin{tabular}{cc}
\small{(a) Facebook Shares (Darker = More Shares) } & \small{(b) Facebook Reactions (Darker = More Reactions) }\\
\includegraphics[width=10.8cm]{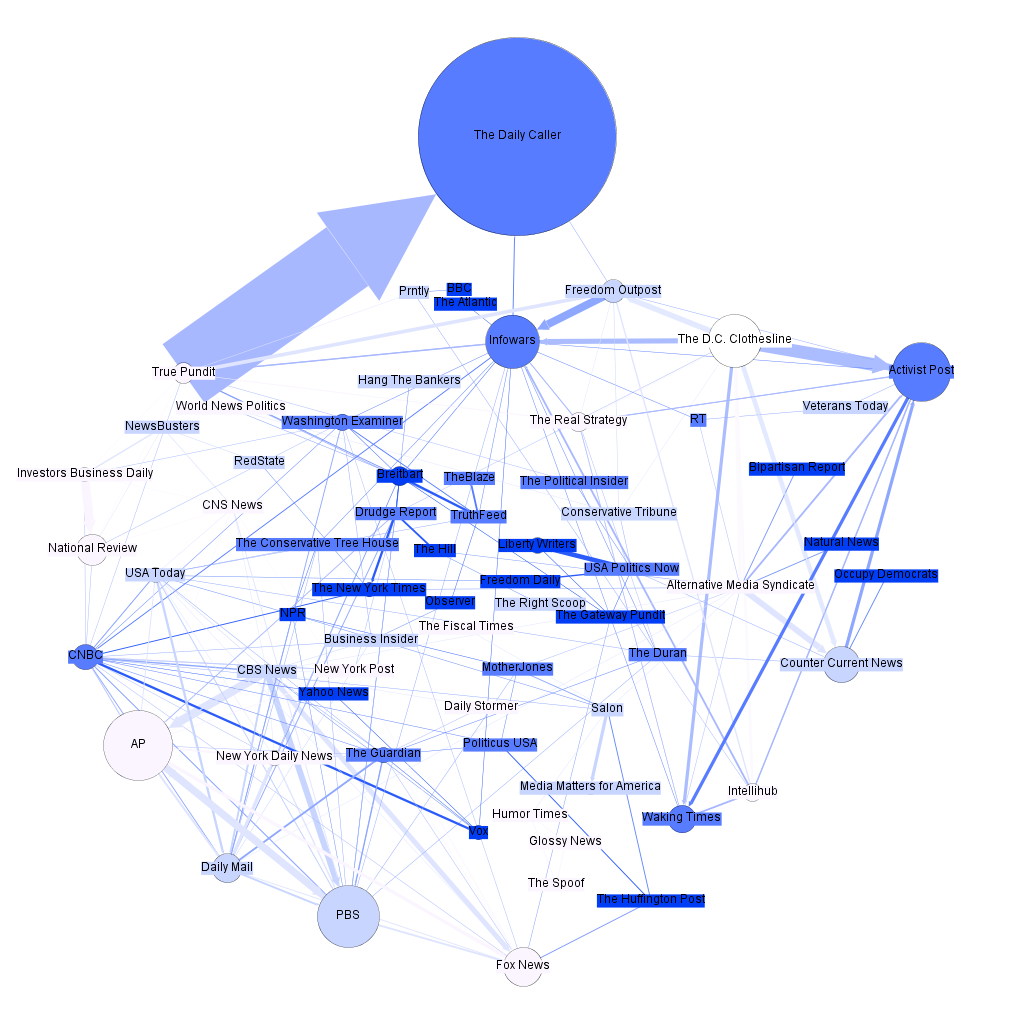}&
\includegraphics[width=10.8cm]{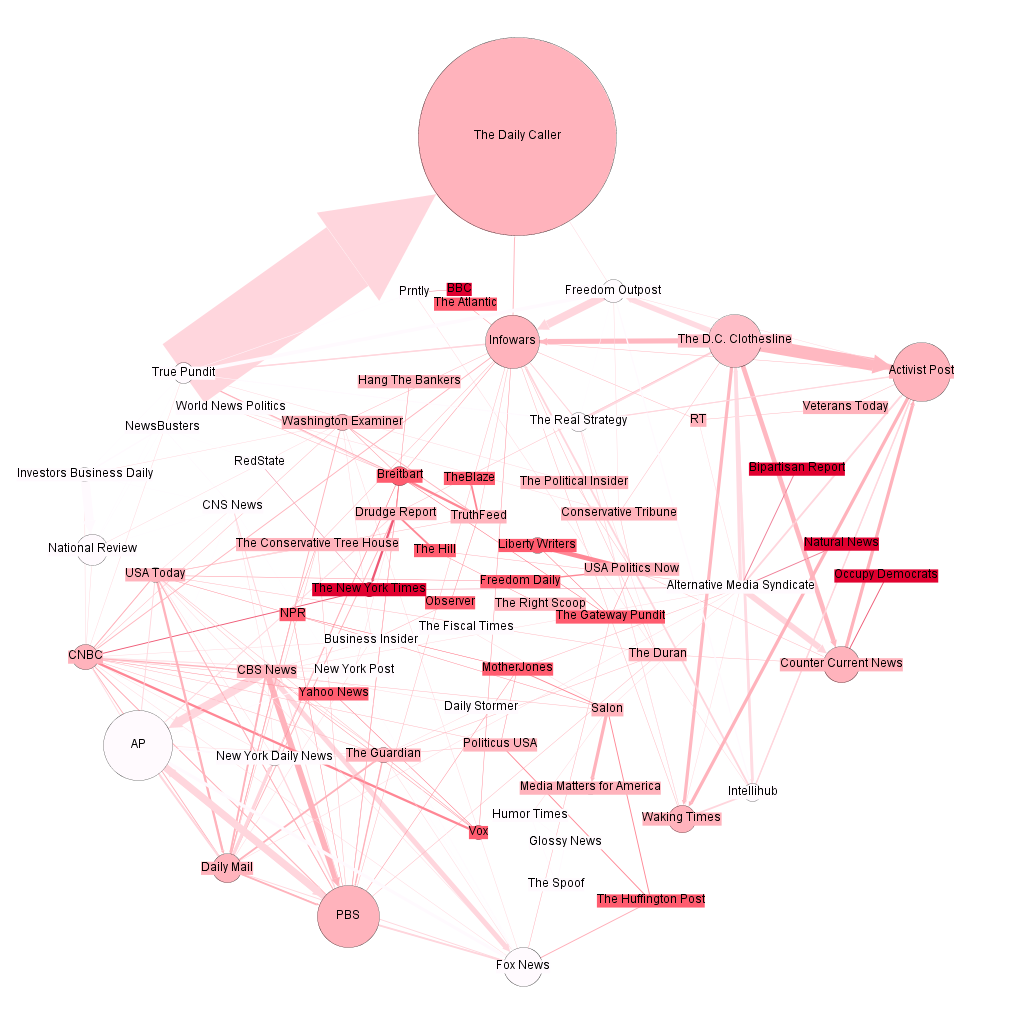}\\
\small{(c) Change Titles By Most (Darker = Changed by More) } & \small{(d) Number of Changed Titles (Darker = More Changed) }\\
\includegraphics[width=10.8cm]{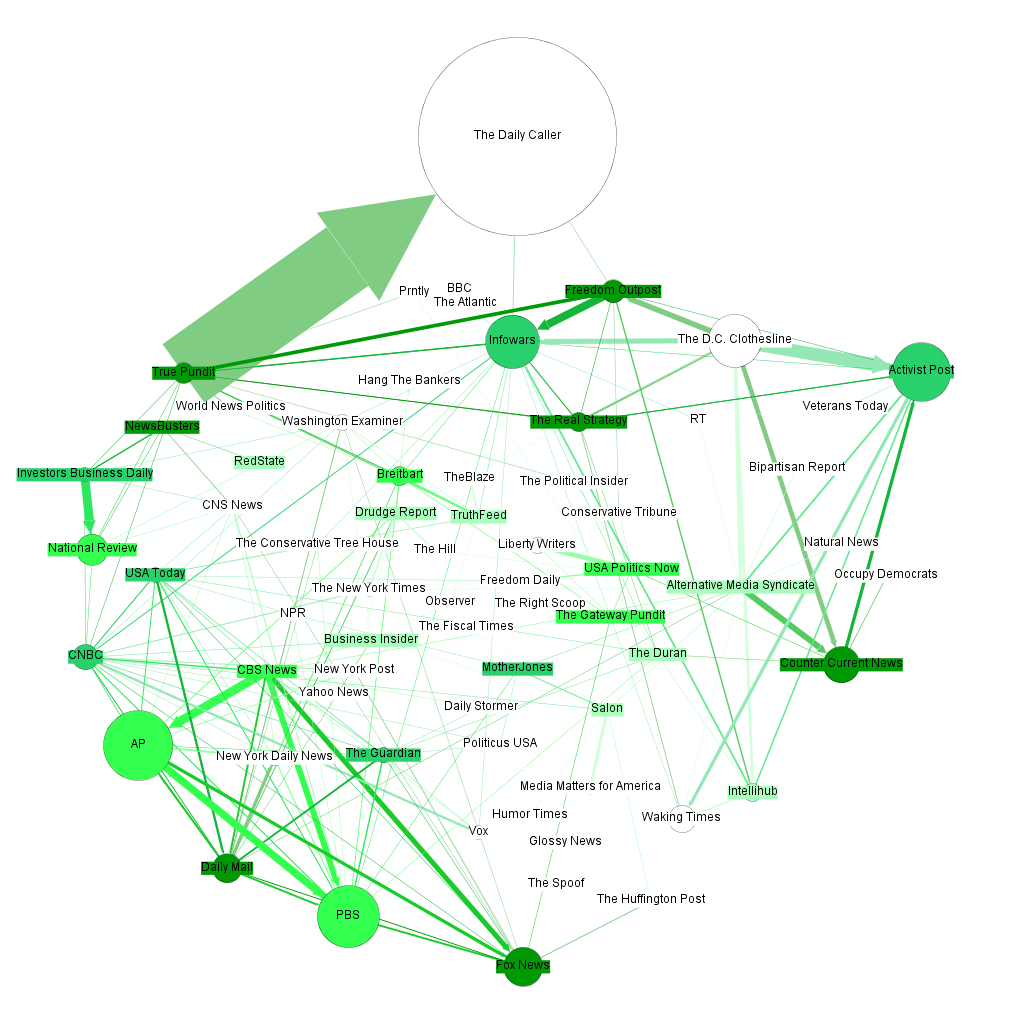}&
\includegraphics[width=10.8cm]{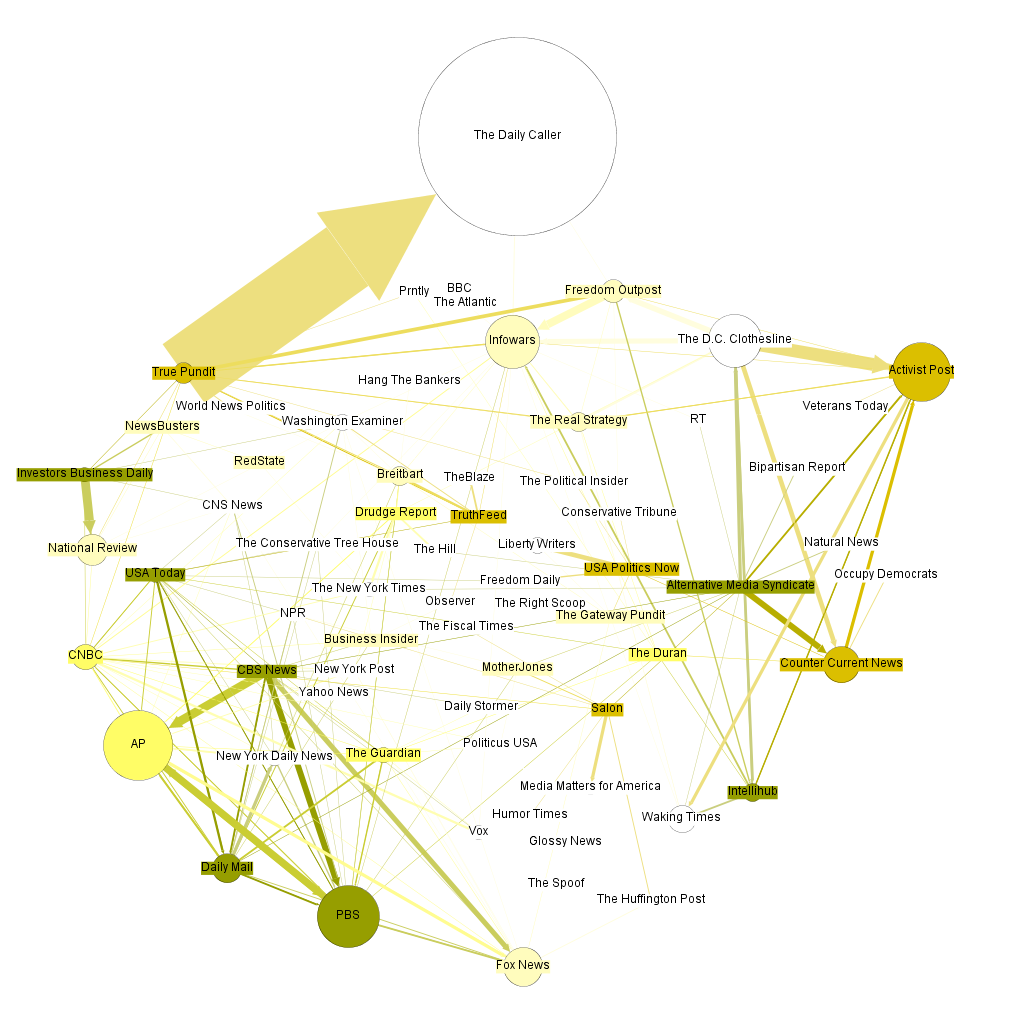}\\
\end{tabular}
 \vspace*{-0.2in}  \caption{Article similarity graphs during all 6 two-week periods. The weighted in-degree is the number of articles copied from a source. The weight is indicated by the size of the arrow. The in-degree of a source is shown by the size of the node.} 
\label{attrib_nets2}
\end{figure*}

Note, not all title changes are bad. For example, when PBS changes titles from AP, the core information does not change and little bias seems to be introduced. While in other cases, the information displayed is completely altered. Thus, we would say a good title change does not misrepresent the article information or create claims not backed by the article. To illustrate the title change in our combined network, we color nodes based on how many titles are changed and how much titles are changed (Figure~\ref{attrib_nets2}c and d). While we see a difference in the top 10 set for these statistics, we see very similar networks in Figures~\ref{attrib_nets2}c and~\ref{attrib_nets2}d.

To better quantify these varying differences, we compute a set of content-based features from~\cite{horne2018accessing} on each title. For the 17 sources found in Figure~\ref{title_top} we compare individual feature distributions. Specifically, we compare feature distributions from each source to feature distributions of all articles they copy using ANOVA hypothesis testing. For example, if we are examining the negative opinion words in the news source Salon, we compute the distribution of negative opinion words in the title of all the Salon articles and the distribution of negative opinion words in the titles of all the articles Salon copied. If those distributions are both normal and significantly different according to ANOVA, this analysis tells us there are features commonly changed in the title by a source. Due to space restrictions, we do not display all of the features computed. For implementation details, please refer to the three studies cited.

In Table~\ref{tab:feat_mostchange}, we show the top features changed based on significance, where significance is a p-value less than 0.05 and the distributions tested are normal and have more than 8 samples. While many more features may be changed at an individual title level, we want to understand what features are consistently and significantly being changed. Over many of the sources in this analysis, we find that more bias and negative opinion words are added. Both the bias words and negative opinion words comes from two lexicons in ~\cite{liu2005opinion} and ~\cite{recasens2013linguistic}. Examples of words in the negative opinion lexicon include: anti-american, disrespectful, and lies. This lexicon contains almost 5K words. Examples of words in the bias word lexicon include: best, corruption, and propaganda. This lexicon includes 654 bias-inducing lemmas. We also find a few sources that consistently increase the number of positive opinion words in the title. Examples of words in the positive opinion lexicon include: accomplished, honest, and improved.

This analysis shows that mainstream sources tend to change less content-based features and more structure based features (punctuation), whereas more alternative news sources tend to change more content based features in the title (bias words, emotion words, etc.).

\begin{table}
    \centering
    \fontsize{8}{10}\selectfont
    \hspace*{-0.1in}\begin{tabular}{|c|c|}\hline
        \textbf{Source} &  \textbf{Significant Features Changed}\\ \hline
        CBS News & less punctuation, slightly more quotes  \\ \hline
        PBS & less punctuation, slightly more quotes\\ \hline
        Breitbart & harder to read, more stopwords and bias words\\ \hline
        The Gateway Pundit & more stopwords, negative words, and bias words\\ \hline
        Alt Media Syn & more stopwords, negative words, and bias words\\ \hline
        The Duran & more stopwords, negative words, and bias words\\ \hline
        Drudge Report & more negative words\\ \hline
        RedState & more negative words and bias words\\ \hline
        Truthfeed & more stopwords, bias words, and negative words\\ \hline
        Business Insider & harder to read\\ \hline
        Intellihub & harder to read, more stopwords\\ \hline
        Salon & more bias words,  more positive words\\ \hline
        Investors Biz Daily & less punctuation\\ \hline
        The D.C. Clothesline & slightly more stopwords\\ \hline
        Daily Mail & more positive words\\ \hline
        USA Today &  more positive words\\ \hline
        Activist Post &  more positive words\\ \hline
    \end{tabular}
    \caption{Most significant features changed when a source changes titles. Significance is determined by ANOVA p-values less than 0.05 on normal distributed features.}
    \label{tab:feat_mostchange}
\end{table}

\section{Conclusion and Future Work}
In this study, we present a first look at content copying across the modern news ecosystem. We find that verbatim content republishing is a fairly common occurrence, with 67 of our 92 sources having at least 1 article republished or copying at least 1 article. We show that there exists clear alternative and mainstream news communities of republishing with sparse connections in between. While sparse, the connections between alternative and mainstream media are consistent through each two-week time-frame. We show that these connections between alternative and mainstream news can be concerning, as the republishing of credible content on unreliable news sites does occur. In general, while the connections between reliable and unreliable news producers exist, it is much more likely that news producers republish material from like-audience news producers (mainstream from mainstream, alternative from alternative, hyper-partisan from hyper-partisan). Looking at specific features changed, we found that mainstream sources tend to change less content-based features and more structure based features, whereas more alternative news sources tend to change more content based features in the title. When alternative sources copy articles from other alternative sources, the titles are rarely changed. While when mainstream sources copy from other mainstream sources, the titles are often changed, but the title intent stays he same. 

In general, we conclude that content copying networks can help label brand-new sources using the content-copying community they belong to. If this type of method can be developed at scale, it can serve as an unsupervised approximation of new source intent. Further, it is possible to use the network to find salient content producers, content broadcasters, and sources that distort credibility information. With these larger goals in mind, this study leaves many directions open for future work. In this study we only explore near verbatim content copying; however, more fine-grained content mixing likely exists. For example, while one disinformation tactic may be to publish completely credible articles on the same page as false articles, it may also be the case that true and false information is mixed in a single article. This article level analysis could provide better insight into malicious news producer's behavior and could make copying network based algorithms more effective.


\bibliographystyle{aaai}
\bibliography{references}

\end{document}